\documentclass[11pt]{article}
\usepackage{amsmath, amssymb}
\usepackage{slashed}
\usepackage{verbatim}
\usepackage{latexsym}
\usepackage{euscript}
\usepackage{amsfonts}
\usepackage{verbatim}
\usepackage[]{array}
\usepackage[]{mathrsfs}
\usepackage{graphicx}
\usepackage{amsmath}
\usepackage{amssymb}
\usepackage{amsxtra}
\usepackage{color}
\def\be{\begin{eqnarray}}
\def\ee{\end{eqnarray}}
\def\0{\nonumber}
\def\d{\partial}
\def\tr{{\rm tr}}

\def\al{\alpha}

\def\det{\rm det}

\def\sfk{{\sf k}}
\def\sfT{{\sf T}}
\usepackage{slashed}
\usepackage{verbatim}
\usepackage{latexsym}\usepackage{color}
\usepackage{euscript}

\newcommand\ET{\EuScript{T}}

\newcommand\EA{\EuScript{A}}

\normalfont\large
\begin{document}
\vskip 2cm
\begin{center}

{\LARGE Elusive anomalies }

\vskip 1cm

{\large  L.~Bonora
\\
\textit{ International School for Advanced Studies (SISSA),\\Via
Bonomea 265, 34136 Trieste, Italy   }
 }
\end{center}

\vskip 1cm
{\bf Abstract}. Usually, in order to compute an anomaly (be it chiral or trace) with a perturbative method, the lowest significant order is sufficient. With the help of gauge or diffeomorphism invariance it uniquely identifies the anomaly. This note is a short review of the ambiguities that arise in the calculation of trace anomalies, and is meant, in particular, to signal cases in which the lowest perturbative order is not enough to unambiguously identify a trace anomaly. This may shed light on some recent contradictory results.

\vskip 1cm

\section{Introduction}

The first manifestation of anomalies in QFT (the Adler-Bell-Jackiw anomaly) originated from an apparently technical problem: the constant shift of an integration variable in a fermion loop integral leads to a vanishing result (which, in turn, implies a conserved chiral current), except for the fact that this integral is UV divergent, so that the shift is illegal; on the contrary, a proper treatment of this problem leads to a non-vanishing result, which, in turn, implies an anomalous conservation law. The way it came up the first time might have seemed to be due to a technicality, but in fact it turned out to be the tip of an iceberg. On the one hand it was the first  of a series of similar results that lead to the discovery of many anomalies: many currents which are classically conserved are not anymore so after quantization. On the other hand these anomalies were derived in a number of ways, both perturbative and nonperturbative, and it was discovered that they are far from wild, random violations of the conservation laws, but, on the contrary, they satisfy group theory motivated consistency conditions. Finally, the illuminating connection was found with the family's index theorem, illuminating, because it revealed that (consistent) anomalies represent obstructions to the existence of the inverse of the Dirac (or Dirac-Weyl) operator, i.e. to the very existence of the fermion propagator. The latter is a fundamental ingredient of a quantum theory, therefore consistent anomalies are a spy of its bad health. 

In quite a similar way, after the ABJ anomalies, also anomalies in the trace of the energy-momentum tensor were found in theories where, classically, conformal invariance requires a traceless e.m. tensor, \cite{capper}. Strangely enough trace anomalies have lived a separate life from chiral anomalies, and any attempt to unify them has failed. Nevertheless it is true that both kind of anomalies are strictly linked to the existence of the inverse kinetic operator: to see it is enough, for instance, to consider that both the derivation of a current conservation law and the tracelessness condition within the path integral approach requires the existence of the inverse kinetic operator. Leaving aside, for the time being, this link between the two types of anomaly (chiral and trace) let us focus now on their differences. 

It is not a mistery that the calculation of trace anomalies has led to some controversial results. The reason is, on the one hand, the ambiguity in the definition of trace anomaly and, on the other hand on the ambiguities intrinsic to their derivation. It was pointed out above that ABJ chiral anomaly arose from resolving an ambiguity in the definition of a loop integral. These types of ambiguities are the basic ones, and, of course, are present also in the case of trace anomalies. But they are not the only ones. The very definition of the trace anomaly in terms of perturbative amplitudes poses a problem. Let us denote by $\langle\!\langle T_{\mu\nu}(x)\rangle\!\rangle$ the full one-loop one-point e.m. tensor, i.e. 
\be
&&\!\!\!\!\!\!\!\!\langle \!\langle T_{\mu\nu}(x)\rangle\!\rangle= \sum_{n=0} ^\infty \frac { i^{n}}{{2^{n}}n!} \int
\prod_{i=1}^n
d^{4} x_i\sqrt{g(x_i)} h^{\mu_i\nu_i}(x_i)\langle 0| {\cal T}T_{\mu\nu}(x) T_{\mu_1\nu_1}(x_1)\ldots T_{\mu_n\nu_n}(x_n)|0\rangle\0\\
&&\label{1ptem}
\ee		
where $\langle 0| {\cal T}T_{\mu\nu}(x) T_{\mu_1\nu_1}(x_1)\ldots T_{\mu_n\nu_n}(x_n)|0\rangle$ are e.m.tensor correlators calculated with Feynman diagrams.		
Then we may proceed in two ways to compute the trace of this object. The first is to evaluate
$ g^{\mu\nu}(x) \langle \!\langle T_{\mu\nu}(x)\rangle\!\rangle$, i.e. to take the trace of the one-loop one-point e.m. tensor. The second is to evaluate $ \langle \!\langle T_\mu^\mu(x)\rangle\!\rangle$, i.e. the one-loop one-point function of the e.m. trace (which is non-vanishing off-shell). In many examples these two quantities are different, therefore we face the problem of defining what we mean by trace anomaly. It turns out that the right definition is the difference between the two
\be
T(x)=g^{\mu\nu}(x) \langle \!\langle T_{\mu\nu}(x)\rangle\!\rangle- \langle \!\langle T_\mu^\mu(x)\rangle\!\rangle\label{Duff}
\ee
proposed by M. J. Duff, \cite{duff1994,duff2020}.
 According to the physical interpretation in \cite{bonora2021}, it is entirely due to the violation of the equation of motion of the theory  (remember that the trace of the e.m. tensor classically incorporates the equation of motion).

But in the Feynman diagram approach there are other ambiguities. When regularizing the loop integral we are faced with more than one possibility, no matter what regularizing prescription we use, dimensional or Pauli-Villars, to name the most frequently used. These possibilities may lead to final results differing by local terms (this may happen also for chiral anomalies). Now, to proceed further, we have to introduce another ingredient that can be, and usually is, disregarded in the case of chiral anomalies: diffeomorphism invariance.The trace anomaly is the response of the functional integral under a rescaling of the metric. Therefore the properties of the metric are inevitably brought into the game, and one has to check in particular that diffeomorphisms are conserved. This requires a recourse to cohomology. As we shall see in an example below, depending on the regularization prescription, the divergence of the e.m. tensor may be vanishing or non-vanishing, giving rise in the latter case to a cocycle of the diffeomorphisms. Such cocycle may be trivial, that is a counterterm can be added to the effective action which cancels this anomaly and, simultaneously, modify the trace anomaly. We shall refer to this as {\it the stabilizing or repairing} role of diffeomorphism conservation. In most situations this is what happens: a unique expression for the trace anomaly is identified, accompanied by conserved diffeomorphisms. In other words, as it should be, the final result does not depend on the regularization prescription. Said another way, a regularization prescription determines the anomaly up to trivial cocycles. 

This is not the end of the story. There is another possible ambiguity which we would like to illustrate in this note. It is rather rare but plays a crucial role in specific cases and renders the lowest order calculation of the trace anomaly unusable. Such an ambiguity is triggered if the  three-point function of the energy-momentum tensor (the lowest order as far as the calculation of the trace anomaly in 4d is concerned) vanishes identically. This may not happen for the full e.m. tensor, but it may happen for its odd-parity part. Since even-parity and odd-parity correlators split neatly we can treat them in the anomaly calculations as separate entities. When such vanishing occurs, the first term in \eqref{Duff} vanishes, but the second need not vanish. On the other hand  the (odd-parity) divergence of the e.m. tensor at the (three-point level) vanishes identically and there is no possibility to adjust the effective action by adding counterterms in such a way as to unambiguously determine the trace anomaly. Now, in most cases the lowest order perturbative calculation is enough to determine trace or gauge anomalies completely, relying on gauge or diffeomorphism invariance, respectively. But in this case it is impossible, the problem is logically undecidable at the lowest perturbative order. The way out is to resort to higher order approximations or to a non-perturbative approach.  

In this short note I would like to present an example of this pathological phenomenon. But, before, in order to appreciate it, it is necessary to understand the repairing mechanism of diffeomorphism conservation. For this reason I present in the next section a simple 2d example in which this mechanism works, and devote the third section to the non-working example. Throughout the paper the reference action will be that of a right-handed Weyl fermion
\be
S= \int d^dx \, \sqrt{g} \, i\overline {\psi_R} \gamma^\mu\left(\partial_\mu
+\frac 12 \omega_\mu \right)\psi_R \label{action}
\ee
where $g=\det( g_{\mu\nu})$, $\gamma^\mu = e^\mu_a \gamma^a$,  ($\mu,\nu,...$ 
are world indices, $a,b,...$ are flat indices)  and $\omega_\mu$ is the spin connection:
\be
\omega_\mu= \omega_\mu^{ab} \Sigma_{ab}\0
\ee
where $\Sigma_{ab} = \frac 14 [\gamma_a,\gamma_b]$ are the Lorentz generators; $\psi_R=
P_R\psi$, where $P_R= \frac {1+\gamma_*}2$, and $\gamma^\ast$ is the appropriate chirality matrix. The reference classical e.m. tensor will be
\be
T_{\mu\nu}= \frac i4 \overline {\psi_R} \gamma_\mu {\stackrel{\leftrightarrow}{\nabla}}_\nu\psi_R+ \{\mu\leftrightarrow \nu\}
 \label{emt0}
\ee
The theory \eqref{action} is invariant under diffeomorphisms $\delta_\xi g_{\mu\nu}=\nabla_\mu \xi_\nu+ \nabla_\nu\xi_\mu$ ($\nabla$ is the gravitational covariant derivative and $\xi_\mu =g_{\mu\nu}\xi^\nu$) and Weyl transformations $\delta_\omega g_{\mu\nu} =2\omega g_{\mu\nu}$, where $\xi^\mu(x)$ and $\omega(x)$ are the relevant local parameters. As a consequence, classically,
\be
\nabla^\mu T_{\mu\nu}(x)=0, \quad\quad T_\mu^\mu(x)=0 \label{classicalconserv}
\ee
on shell.

\section{A simple (working) example}

In two dimensions, due to the anticommutativity of spinors, the spin connection drops out of the action \eqref{action}. For $\psi_R$ the action becomes
\be
S= i\int d^2x \, \sqrt{g} \, \overline {\psi}_R \gamma^\mu \partial_\mu
\psi_R\label{SR}
\ee
Although this is not strictly necessary, we will further simplify it by absorbing the $\sqrt{g}$ into a redefinition of $\psi$: $\psi\to \widetilde \psi = g^{\frac 14} \psi$. 
\be
\widetilde S= i\int d^2x \,  {\overline {\widetilde\psi}}_R \gamma^\mu
\partial_\mu \widetilde\psi_R= i\int d^2x \,  {\overline {\widetilde\psi}}_R
\gamma^a e_a^\mu \partial_\mu \widetilde\psi_R\label{SLtilde}
\ee

Now let us write $e_a^\mu\approx \delta_a^\mu-\chi_a^\mu$ and make the
identification $2\chi_a^\mu=h_a^\mu$,  where $h_{\mu\nu}$ is the
gravitational fluctuation field: $g_{\mu\nu}=\eta_{\mu\nu}+h_{\mu\nu}$. 
The fermion propagator is
\be 
\frac i{\slash \!\!\! p+i\epsilon }\label{prop}
\ee
and there is only one graviton-fermion-fermion ($V_{ffg}$) vertex
given by
\begin{equation}
\dfrac{i}{8}\left[
\left(p+p'\right)_\mu\gamma_\nu+\left(p+p'\right)_\nu\gamma_\mu\right]\frac{
1+\gamma_*}{2}.\label{FRule1}
\end{equation}
where $p$ and $p'$ are the two graviton momenta, one entering the other exiting. Other vertices will not be relevant.

Our purpose is to compute the two terms in eq.\eqref{Duff}. In 2d the lowest order contribution is given by the two-point amplitudes
\be
\eta^{\mu\nu} \langle 0| {\cal T}T_{\mu\nu}(x)T_{\lambda\rho}(y)|0\rangle \quad\quad
{\rm and}\quad\quad  \langle 0| {\cal T}T_\mu^\mu(x)T_{\lambda\rho}(y)|0\rangle, \label{twoterms}
\ee
respectively. 
Their Fourier transforms is provided by a Feynman (bubble) diagram with a fermion loop with momentum $p$ and two external gravitons (one entering, one exiting) with momentum $k$. More in detail, considering the first term in \eqref{twoterms} we have 
\be
\langle T_{\mu\nu}(x) T_{\lambda\rho}(y)\rangle =4\int
\frac{d^2k}{(2\pi)^2}e^{-ik(x-y)}\widetilde 
\ET_{\mu\nu\lambda\rho}(k)\label{TmnTlr2d}
\ee
with
\be
\widetilde \ET_{\mu\nu\lambda\rho}(k)&=&-\frac 1{64} \int  
\frac{d^2p}{(2\pi)^2} 
{\rm tr}
\left(\frac 1 {\slashed{p}} (2p-k)_\mu\gamma_\nu \frac
{1+\gamma_\ast}2\frac
1{\slashed{p}-\slashed{k}} (2p-k)_\lambda \gamma_\rho \frac
{1+\gamma_\ast}2\right)+
\left\{ \begin{array}{c}\mu\leftrightarrow \nu \\ \lambda \leftrightarrow
\rho\end{array} \right\}.\label{T(k)2d}
\ee
The last bracket means that we have to add three more terms like the first so as to realize a symmetry under the exchanges $\mu \leftrightarrow \nu$, $\lambda \leftrightarrow \rho$. Moreover, we have to symmetrize  with respect to the exchange $(\mu,\nu) 
\leftrightarrow (\lambda,\rho)$ (bosonic symmetry). 

The integral in \eqref{T(k)2d} is UV divergent. We proceed to regularize it with dimensional regularization.To this end, as usual, we introduce extra 
space components of the momentum running around the loop, $p_\mu\to p_\mu+\ell_{\bar \mu}$ 
($\ell_{\bar \mu}=\ell_2,\ldots,\ell_{\delta+1}$) . So  \eqref{T(k)2d} becomes
\be
\widetilde \ET^{(reg)}_{\mu\nu\lambda\rho}(k)&=&-\frac 1{64} \int  \frac{d^2p\, 
d^\delta\ell}{(2\pi)^{2+\delta}} {\rm tr}
\Big(\frac 1 {\slashed{p}+{\slashed{\ell}}} (2p-k)_\mu\gamma_\nu \frac
{1+\gamma_\ast}2\frac
1{\slashed{p}+\slashed{\ell}-\slashed{k}} (2p-k)_\lambda\gamma_\rho \frac
{1+\gamma_\ast}2\Big).\label{T(k)2dreg1}
\ee
Now let us recall that $\slashed {p}^2=p^2, \slashed {\ell}^2=-\ell^2$, 
$\slashed {p}\slashed {\ell}+\slashed {\ell}\slashed {p}=0$ and $[\gamma_\ast, 
\slashed {\ell}]=0$, while
$\{\gamma_\ast, \slashed {p}\}=0$. Moreover 
$\tr{\left(\gamma_{\mu}\gamma_{\nu}\gamma_{*}\right)}=- 2^{1+\frac {\delta}2} \epsilon_{\mu\nu}$. 
Working out the $\gamma$-matrix algebra and performing a Wick rotation $k_0\to ik_0$ one can compute the loop integral.  Here, for simplicity, we report only the even parity part of the trace and the divergence of the e.m. tensor:
\be
\widetilde \ET^{E\mu}_{\mu\lambda\rho}(\sfk)=\frac i{192\pi} \Big[ \sfk_\lambda 
\sfk_\rho + \sfk^2 \eta_{\lambda\rho}\Big]\label{traceETE}
\ee
and
\be
\sfk^\mu \widetilde \ET^E_{\mu\nu\lambda\rho}(\sfk)&=& - \frac 
i{384\pi}\Big[\sfk_\nu\sfk_\lambda\sfk_\rho+ \frac 12 
\sfk^2\bigl(\eta_{\nu\lambda}\sfk_\rho + \eta_{\nu\rho} \sfk_\lambda\bigr)\Big]\label{kmuETE}
\ee
where ${\sfk}_\mu$ denotes the Euclidean momentum (in particular $\sfk^2=-k^2$). Next one anti-Fourier-transforms these amplitudes and, after returning to the Minkowski background, inserts them into \eqref {1ptem}. To obtain the corresponding integrated cocycles, one multiplies the first by $\omega$ and saturates the second with $\xi^\nu$ and integrate over space-time. The result are
the two cocycles
\be
\Delta_\omega &=&\!\! \frac 12\!\! \int\! d^2x \, \omega(x)\!\! \int d^2y\, h^{\lambda\rho}(y) 
\langle 0|{\cal T}T_\mu^\mu(x) T_{\lambda\rho}(y) |0\rangle_c 
\label{Deltaomega}\\
&=&\!\! 2 \!\!\int\! d^2x \, \omega(x) \!\!\int\! d^2y\, h^{\lambda\rho}(y) \int \frac 
{d^2k}{(2\pi)^2}\,e^{-ik\cdot(x-y)} \widetilde \ET^\mu_{\mu\lambda\rho}(k)= \frac 1{96\pi} \int d^2x \, \omega(x) \left[ \partial_\lambda\partial_\rho 
h^{\lambda\rho}(x)-\square h_\lambda^\lambda (x)\right]\0
\ee
and
 \be
\Delta_\xi &=&-\frac 12 \int d^2x \, \xi^\nu(x)  \int d^2y\, h^{\lambda\rho}(y) 
(-ik^\mu)e^{-ik\cdot(x-y)}
\widetilde \ET_{\mu\nu\lambda\rho}(k)\0\\
&=&\frac 1{192\pi} \int d^2x \,  \xi^\nu(x)  
\Big[\partial_\nu\partial_\lambda\partial_\rho h^{\lambda\rho}(x) - 
\partial_\lambda \square h_\nu^\lambda(x) \Big]\label{Deltaxi}
\ee
Let us recall that the parameters $\xi^\mu$ and $\omega$, according to the rule of BRST quantization, are promoted to anti-commuting fields. The lowest order transformation rules for  the metric is $\delta_\omega^{(0)} h_{\mu\nu}= 2\omega \eta_{\mu\nu}$ and $\delta_\xi h_{\mu\nu}=\partial_\mu \xi_\nu+\partial_\nu\xi_\mu$. Using this it is easy to prove that
\be
\delta^{(0)}_\omega \Delta_\omega=0, \quad\quad \delta^{(0)}_\xi \Delta_\xi=0, 
\quad\quad
\delta^{(0)}_\omega \Delta_\xi+\delta^{(0)}_\xi 
\Delta_\omega=0.\label{CComegaxi}
\ee
Both trace and diffeomorphism cocycles are non-vanishing. However the diffeomorphism one is trivial. For let us consider the local counterterm
\be
{\cal C}^{(even)}= \frac 1{384\pi} \int d^2x\, \bigl(h_\rho^\nu(x) 
\partial_\lambda\partial_\nu h^{\lambda \rho}(x) -h_{\lambda \rho}(x)\square 
h^{\lambda \rho}(x)\bigr)\label{countCeven}
\ee
It is easy to prove that
\be
\Delta^{'(even)}_\xi\equiv \Delta^{(even)}_\xi + \delta^{(0)}_\xi {\cal 
C}^{(even)}=0\label{EAxi}
\ee
Therefore diffeomorphisms are conserved. On the other hand the overall even trace cocycle becomes
\be
\Delta^{'(even)}_\omega\equiv \Delta^{(even)}_\omega+ \delta^{(0)}_\omega {\cal 
C}^{(even)}
= \frac 1{48\pi} \int d^2x \, \omega\left[ \partial_\lambda\partial_\rho 
h^{\lambda\rho}-\square h_\lambda^\lambda\right]\label{Deltaprimeomega}
\ee
So far we have computed the first term of \eqref{Duff}. We have to compute also the second, i.e. the second one in \eqref{twoterms}.  The corresponding amplitude, once regulated, takes the form
\be
\widetilde {\widehat\ET}{}^\mu{}_{\mu \lambda\rho }(k) = -\frac 1 {64} \int \frac
{d^2p\, d^{\delta}\ell}{(2\pi)^{2+\delta}} \tr\left(
\frac{\slashed{p}+\slashed{\ell} }{p^2-\ell^2} \, \left(2\slashed{p}+ 2 \slashed{\ell} -\slashed{k}\right) \frac {\slashed{p}-\slashed{k}}{(p-k)^2-\ell^2}(2p-k)_\lambda
\gamma_\rho\frac {1+\gamma_\ast}2 \right)\label{T'(k)2dreg}
 \ee
A direct calculation shows that it vanishes. Therefore the trace anomaly is determined by \eqref{Deltaprimeomega}, which is the first order approximation of
\be
\EA^{(even)}_\omega=   \frac 1{48\pi} \int d^2x \,\sqrt{g}\, \omega \, R\label{EAomegaR}
\ee

\subsection{Another prescription}

The regularizing prescription \eqref{T(k)2dreg1} is not the only possibility. We could have started from
\be
\widetilde \ET'_{\mu\nu\lambda\rho}(k)&=&-\frac 1{64} \int  
\frac{d^2p}{(2\pi)^2} 
{\rm tr}
\left(\frac 1 {\slashed{p}} (2p-k)_\mu\gamma_\nu \frac
1{\slashed{p}-\slashed{k}} (2p-k)_\lambda \gamma_\rho \frac
{1+\gamma_\ast}2\right).\label{T(k)2d'}
\ee
and regularize it as follows
\be
\widetilde \ET^{(reg)'}_{\mu\nu\lambda\rho}(k)=-\frac 1{32} \int  \frac{d^2p\, 
d^\delta\ell}{(2\pi)^{2+\delta}} 
{\rm tr}
\Big(\frac 1 {\slashed{p}+{\slashed{\ell}}} (2p-k)_\mu\gamma_\nu\frac
1{\slashed{p}+\slashed{\ell}-\slashed{k}}  (2p-k)_\lambda\gamma_\rho \frac
{1+\gamma_\ast}2\Big).\label{T(k)2dreg1'}
\ee
We shall refer to it is the rightmost $\gamma_\ast$ prescription.  The result is now
\be
\widetilde \ET^{'E\mu}_{\mu\lambda\rho}(\sfk)=\frac i{96\pi} \Big[ \sfk_\lambda 
\sfk_\rho + \sfk^2 \eta_{\lambda\rho}\Big]\label{traceETE'}
\ee
and
\be
\sfk^\mu \widetilde \ET^E_{\mu\nu\lambda\rho}(\sfk)&=&0\label{kmuETE'}
\ee
Contrary to the previous prescription this one yields diffeomorphism invariance and the same trace cocycle \eqref{Deltaprimeomega}. It remains for us to evaluate the second term of \eqref{Duff}. The corresponding regulated amplitude is 
\be
{\widetilde {\widehat\ET}}{}'{}^\mu{}_{\mu \lambda\rho }(k)  \stackrel{reg}{=} -\frac 1 {32} \int 
\frac
{d^2p \,d^{\delta}\ell}{(2\pi)^{2+\delta}}\tr\left(
\frac{\slashed{p}+\slashed{\ell}}{p^2-\ell^2} \, 
\left(2\slashed{p}+2\slashed{\ell} -\slashed{k}\right) 
\frac {\slashed{p}+\slashed{\ell}-\slashed{k}}{(p-k)^2-\ell^2}(2p-k)_\lambda
\gamma_\rho
\frac {1+\gamma_\ast}2 \right)\label{ETtrace'}
 \ee
which, again, vanishes. Therefore the two prescriptions lead, as expected, to the same result, the trace anomaly \eqref{EAomegaR}, while diffeomorphisms are conserved (as far as the even parity part is concerned.). 

In this section we have illustrated an example (probably the simplest one) of a perfectly working cohomological mechanism: the first prescription leads both to a trace and a diffeomorphism anomaly; however the latter is trivial and can be eliminated with a counterterm, which in turn modifies the trace anomaly giving it the final (minimal) form. The second prescription preserves diffeomorphism invariance and yields the previous final form of the trace anomaly. In the next section we shall see an example in which this mechanism cannot work.

\section{The problematic example}

The example we consider in the sequel is that of a right-handed Weyl fermion coupled to a non-trivial metric. The action is \eqref{action} with $d=4$, but in this case the spin connection does not drop out. One can write the action as follows
\be 
S= \int d^4x \, \sqrt{|g|} \,\left[ \frac i2\overline {\psi_R} \gamma^\mu {\stackrel{\leftrightarrow}{\d}}_\mu  \psi_R -\frac 14\epsilon^{\mu a b c} \omega_{\mu a b} \overline{\psi_R} \gamma_c \gamma_5\psi_R\right]
\label{actionR1}
\ee
where it is understood that the derivative applies to $\psi_R$ and $\overline {\psi_R}$ only. We have used the relation $\{\gamma^a, \Sigma^{bc}\}=i \, \epsilon^{abcd}\gamma_d\gamma_5$.
We expand $g_{\mu\nu}$ and $e^a_\mu$ as before, and, accordingly 
\be 
\omega_{\mu a b}\, \epsilon^{\mu a b c}= - \epsilon^{\mu a b c}\,  \d_\mu \chi_{a\lambda}\,\chi_b^\lambda+...\label{omega}
\ee
Proceeding as in the previous 2d example, after some algebra the action takes the form
\be 
S\approx \int d^4x \,  \left[\frac i2 (\delta^\mu_a -\chi^\mu_a ) \overline {\psi_L} \gamma^a {\stackrel{\leftrightarrow}{\d}}_\mu  \psi_L +\frac 14\epsilon^{\mu a b c}\,  \d_\mu \chi_{a\lambda}\,\chi_b^\lambda\,  \overline\psi_L \gamma_c \gamma_5\psi_L\right]\0
\ee
from which we can extract the Feynman rules. The fermion propagator and fermion-fermion-graviton vertex ($V_{ffg}$) are the same as before. In addition we have a two-fermion-two-graviton vertex
 ($V_{ffgg}$)
\be 
\frac 1{64} t_{\mu\nu\mu'\nu'\kappa\lambda}(k-k')^\lambda\gamma^\kappa\frac {1+\gamma_5}2\label{2f2g}
\ee 
where
\be
t_{\mu\nu\mu'\nu'\kappa\lambda}=\eta_{\mu\mu'} \epsilon_{\nu\nu'\kappa\lambda} +\eta_{\nu\nu'} \epsilon_{\mu\mu'\kappa\lambda} +\eta_{\mu\nu'} \epsilon_{\nu\mu'\kappa\lambda} +\eta_{\nu\mu'} \epsilon_{\mu\nu'\kappa\lambda}\label{t}
\ee
and the graviton momenta $k,k'$ are incoming. Other vertices are irrelevant for the sequel.

This model has no even- nor odd-parity diffeomorphism anomalies, while it has both even and, as we shall see, odd-parity trace anomalies. The even part works much in the same way as in the previous 2d example. Our interest in this section is focused on the odd parity part. It is well-known that in 4d there cannot be odd-parity consistent diffeomorphism anomalies, but a priori we cannot exclude other (trivial) anomalies related to the trace anomalies (much like the \eqref{Deltaxi} above). Therefore we have to verify that odd-parity divergence of the e.m. tensor vanishes. The relevant lowest order contribution (which we denote symbolically by $\langle \partial \!\cdot \! T T T\rangle$) may come from a triangle and a bubble diagram. The triangle diagram contains three  fermion propagators and three $V_{ffg}$ vertices. Taught by the 2d example we use the rightmost $\gamma_\ast\equiv \gamma_5$ prescription. The corresponding Fourier-transformed contribution after regularization is 
\be 
&& \!\!\!\!q^\mu\widetilde  \ET^{(odd)}_{\mu\nu\lambda\rho\alpha\beta}(k_1,k_2) =-\frac 1 {512}\int
\frac
{d^4pd^\delta\ell}{(2\pi)^{4+\delta}}\, {\rm tr} \left[\left(\frac
{\slashed{p}+\slashed{\ell}}{p^2-\ell^2}(2p-k_1)_\lambda
\gamma_\rho\right.\right.\frac
{\slashed{p}-\slashed{k}_1+\slashed{\ell}}{(p-k_1)^2-\ell^2}\0\\
&& \times (2p-2 k_1 -k_2)_{\al}\gamma_{\beta}
\left.\left.\frac {\slashed{p} - \slashed{q}+\slashed{\ell}}{(p-q)^2-\ell^2}\bigl( (2{p}
-{q})\cdot q\,\gamma_\nu- (2{p}-{q})_\nu \slashed{q}\bigr)\right) \,
\gamma_5\right] \0
\ee
It is not difficult to show that it vanishes identically. Also the contribution from the bubble diagram, constructed with one $V_{ffg}$, one  $V_{ffgg}$ and two propagators, vanishes.  Therefore we conclude that with this prescription diffeomorphisms are exactly preserved. 

We  next compute the odd parity contribution of the triangle and bubble diagram to the trace anomaly. At first sight this calculation does not seem to make sense, because a well known result of CFT claims that a conformal odd-parity three-point function $\langle 0|{\cal T}T_{\mu\nu}(x) T_{\mu'\nu'}(y) T_{\alpha\beta}(z)|0\rangle^{(odd)}$  vanishes identically for algebraic reasons.
This is confirmed by a direct calculation. In fact one can prove that, with both prescriptions above, 
$\langle 0|{\cal T}T_{\mu\nu}(x) T_{\mu'\nu'}(y) T_{\alpha\beta}(z)|0\rangle^{(odd)}$ vanishes. So, at the lowest significant perturbative order, we can write: 
\be
\eta^{\mu\nu}\langle\! \langle T_{\mu\nu}(x)\rangle\!\rangle^{(odd)} =0 \label{Zhiboedov}
\ee

However according to the definition \eqref{Duff} we must compute also the second term with one insertion of a trace of the e.m. tensor (which we denote by $\langle t T T\rangle$). The triangle diagram gives
 \begin{eqnarray}
&&\widetilde{\sf T}_{\mu\nu\mu'\nu'}(k_1,k_2)\!\!=\!\!  -\frac{1}{256}\int \frac{d^4p}{(2\pi)^4} \int
\frac{d^{ \delta}\ell}{(2\pi)^{\delta}}\mathrm{Tr}\left\{\frac{\slashed{p}
+\slashed{\ell}}{p^2-\ell^2}\left[(2p-k_1)_\mu\gamma_\nu +
(\mu \leftrightarrow
\nu)\right]\frac{(\slashed{p}+\slashed{\ell}-\slashed{k}_1)}{(p-k_1)^2-\ell^2}\right.\0\\
&\times&\!\!\!\!\!\!\left.\left[
(2p-2k_1-k_2)_{\mu'}\gamma_{\nu'}+(\mu' \leftrightarrow
\nu')\right]
{\frac{(\slashed{p}+\slashed{\ell}-\slashed{k}_1-\slashed
{k}_2)}{(p-k_1-k_2)^2-\ell^2}(2\slashed{p}+2\slashed{\ell}-\slashed{k}
_1-\slashed{k}_2)}\left(\frac{
1+\gamma_5}{2}\right)\right\}\,.\label{triangle4}
\end{eqnarray}
which, with the addition of the cross diagram, leads to the following result
\begin{equation}
\widetilde{ \sfT}_{\mu\nu\mu'\nu'}(k_1,k_2) = \frac{1}{6144\pi^2}k^{\alpha}_1
k^{\beta}_2\left(\left(k_1^2+k_2^2+k_1\! \cdot\! k_2\right)t_{\mu\nu\mu'\nu'\alpha\beta}
-t^{(21)}_{\mu\nu\mu'\nu'\alpha\beta}\right)\,,
\label{triangle24}
\end{equation}
where 
\be 
t^{(21)}_{\mu\nu\mu'\nu'\kappa\lambda}=k_{2\mu}k_{1\mu'} \epsilon_{\nu\nu'\kappa\lambda} + k_{2\nu}k_{1\nu'}\epsilon_{\mu\mu'\kappa\lambda} +k_{2\mu}k_{1\nu'} \epsilon_{\nu\mu'\kappa\lambda} +k_{2\nu}k_{1\mu'} \epsilon_{\mu\nu'\kappa\lambda}\label{t21}
\ee
The contribution from the bubble diagram vanishes.

The conclusion of this computation is that the contribution to the odd-parity trace anomaly according to formula \eqref{Duff} comes solely from \eqref{triangle24}.

To simplify the derivation we set the external lines on-shell, $k_1^2=k_2^2=0$. This requires a comment.

\subsubsection{On shell conditions}

Putting the external lines on shell means that the corresponding fields have to satisfy the EOM of Einstein-Hilbert gravity $R_{\mu\nu}=0$.  In the linearized form this means
\be 
\square \chi_{\mu\nu}= \d_{\mu} \d_{\lambda}\chi^\lambda _\nu + \d_\nu \d_{\lambda}\chi^\lambda _\mu -\d_\mu\d_\nu \chi^\lambda_\lambda=0\label{linrmunu}
\ee
We also choose the De Donder gauge: $\Gamma_{\mu\nu}^\lambda g^{\mu\nu}=0$,
which at the linearized level becomes $2\d_\mu\chi^{\mu}_\lambda -\d_\lambda \chi^\mu_\mu=0$. In this gauge (\ref{linrmunu}) becomes
\be 
\square \chi_{\mu\nu}=0\label{KG}
\ee
In momentum space this implies that $k_1^2=k_2^2=0$. We remark that this does not trivially disrupt the cohomology, but define a restricted cohomology of the diffeomorphisms and the Weyl transformations: the latter is defined up to terms $\square  h_{\mu\nu}$ and $\square \xi^\mu$. This is a well defined cohomology, under which we have, in particular,
\be
\delta_\xi \left(2\d_\mu\chi^{\mu}_\lambda -\d_\lambda \chi^\mu_\mu\right) = 2\ \square \xi_\lambda \approx 0\label{Dedonder}
\ee
i.e. in this restricted cohomology the De Donder gauge fixing is irrelevant. Similarly, the term corresponding in momentum space to $k^{\alpha}_1 k^{\beta}_2\, (k_1^2+k_2^2)t_{\mu\nu\mu'\nu'\alpha\beta}$ remains null after a restricted diffeomorphism transformation. The restricted cohomology has the same odd class (the Pontryagin one) as the unrestricted one, i.e. it completely determines it (this is not true for the even classes). Since we know that the final result must be covariant and that there is no covariant extension to all order of the term $k^{\alpha}_1 k^{\beta}_2\, (k_1^2+k_2^2)t_{\mu\nu\mu'\nu'\alpha\beta}$,  the simplification of considering it null does not jeopardize it. This means that this term must be trivial in some way. We will comment on this below.

\subsubsection{Overall contribution}
\label{ssec:overallcontr}

The overall one-loop contribution to the trace anomaly in momentum space, {\it as far as the parity violating part is concerned}, is given by (\ref{triangle24}). After returning to the Minkowski metric 
and Fourier-antitransforming it, we can extract the local expression of the trace anomaly, by replacing the results found so far in (\ref{1ptem}). The result, to lowest order, is 
\be 
\langle\!\langle T^{\mu}_{\mu}(x)\rangle\!\rangle^{(odd)} \approx \frac i{768\pi^2}\epsilon^{\mu\nu\lambda \rho} \left(\d_\mu\d_\sigma h^\tau_\nu \, \d_\lambda\d_\tau h_{\rho}^\sigma-\d_\mu\d_\sigma h^\tau_\nu \, \d_\lambda\d^\sigma h_{\tau\rho}\right)\label{final1}
\ee
 Since
\be
\epsilon^{\mu\nu\lambda \rho}R_{\mu\nu}{}^{\sigma\tau}R_{\lambda\rho\sigma\tau}=
 \epsilon^{\mu\nu\lambda \rho} \left(\d_\mu\d_\sigma \chi^a_\nu \, \d_\lambda\d_a \chi_{\rho}^\sigma-\d_\mu\d_\sigma \chi^a_\nu \, \d_\lambda\d^\sigma \chi_{a\rho}\right)+...\label{epsRR}
\ee
we obtain
\be 
 \langle\! \langle T^{\mu}_{\mu}(x)\rangle\!\rangle^{(odd)} = \frac i{768\pi^2} \, \frac 12\,\epsilon^{\mu\nu\lambda \rho}R_{\mu\nu}{}^{\sigma\tau}R_{\lambda\rho\sigma\tau}\label{final2}
\ee
Now applying the definion \eqref{Duff} and recalling \eqref{Zhiboedov},
we obtain the covariant expression of the parity violating part of the trace anomaly for a Weyl fermion 
\be
T[g](x)=   \frac i{768\pi^2} \, \frac 12\,\epsilon^{\mu\nu\lambda \rho}R_{\mu\nu}{}^{\sigma\tau}R_{\lambda\rho\sigma\tau}.\label{Pontryagintrace}
\ee

\subsection{The missing mechanism}

The trace anomaly \eqref{Pontryagintrace} coincides (up to a coefficient) with the KDS (Kimura-Delbourgo-Salam) anomaly of the chiral current in a theory of Dirac fermions immersed in a non-trivial metric background. In \cite{bonora2015} this coincidence has been explained. Therefore, is everything ok? No, because the term  $k^{\alpha}_1 k^{\beta}_2\, (k_1^2+k_2^2)t_{\mu\nu\mu'\nu'\alpha\beta}$ we have disregarded has not been explained yet, and the attempt to explain it reveals an unexpected obstacle. It corresponds to an integrated anomalous term $\sim \int d^4x\,\omega\, \epsilon^{\mu\nu\lambda\rho} \partial_\mu\square h_\nu^\alpha \partial^\lambda h_{\mu\alpha}$. There is no covariant expression that to the lowest order has this form. Therefore it must be a trivial term. Such a lowest order cocycle can be canceled by a counterterm $\sim \int d^4x h_\mu^\mu \epsilon^{\mu\nu\lambda\rho} \partial_\mu\square h_\nu^\alpha \partial^\lambda h_{\mu\alpha}$. But this counterterm term destroys diffeomorphism invariance. There seems to be no way out.

Before surrendering, one may try to change the regularization prescription. For instance we might use the first prescription of the previous section. It is easy to see that with this new prescription diffeomorphisms are still conserved, as one can directly verify (and as it should be, because of the general theorem in \cite{zhiboedov}). But the trace anomaly changes, both in its form and in its overall coefficient (even the bubble diagram gives a nontrivial contribution). This is a puzzle. We have seen above an example, but many others can be envisaged, where, after some calculations, nonzero trace anomalies and nonzero diffeomorphism anomalies appear in couples, and (unless the the diffeomorphism anomaly is non-trivial, which is not possible in 4d) by subtracting a counterterm we can recover diffeomorphism invariance and modify the trace anomaly to a minimal form. In other words diffeomorphism invariance plays a `repairing' or `stabilizing' role in cohomology. I.e. the diffeomorphism cohomology accompanies the regularization scheme in such a way that the latter preserves the cohomology class. However a necessary condition for this role to be effective is that the relevant amplitude be non-vanishing. Which is not what happens in our puzzling case. {\it In fact, the true origin of the puzzle is not the regularization scheme, but the accidental vanishing of the odd three-point function of the e.m. tensor.} 

The next question is: does that mean that the perturbative calculation of the trace anomaly is impossible? The answer is: no, it is only more difficult. In our particular case the problem arises from the vanishing of the odd three-point function of the e.m. tensor. However the three-point function corresponds to the lowest possible order yielding a significant contribution to the calculation of the trace anomaly. But of course one should consider also the four-point function, the five-point function, and so on. In general there is no such accidental vanishing for the higher order functions. Therefore we should calculate, for instance, the odd four-point function of the energy-momentum tensor and compute both the trace and the divergence in the same way we have done for the three-point function. In this way the stabilizing effect of diffeomorphisms (together with the possible contribution of other graphs, such as the bubble one) would unfold undisturbed. The trouble here is the technical complexity. There is an easier way: a non-perturbative approach.  Appropriate non-perturbative methods exist, they are the Seeley-Schwinger-DeWitt or heat kernel methods: the diffeomorphism invariance is inbuilt in them and, being non-perturbative, they encompass all the relevant higher order amplitudes. The relevant calculations have been carried out in \cite{bonora2018} and more recently in \cite{Liu2022} using a method \`a la Fujikawa. The two calculations lead to the same result, eq.\eqref{Pontryagintrace}, which is also in accord with the general formulas of \cite{duff2}.
\vskip 0.5cm
{\bf Remark} It is worth pointing out that a missing contribution from the perturbative calculation of an anomaly, such as the one we have come across above, is not rare. Let us consider, for instance, the (multiplet) non-Abelian covariant anomaly  $\sim\epsilon^{\mu\nu\lambda\rho}\tr ( T^a F_{\mu\nu} F_{\lambda\rho})$, which appears in the divergence of the chiral current in a theory of Dirac fermions coupled to a  vector potential $V_\mu= V_\mu^a T ^a$  (with curvature $F_{\mu\nu}$). This anomaly contains a quartic term in the potential $V_\mu=V_\mu^a T^a$, which can come only from a pentagon diagram. This diagram however is UV convergent. Therefore the quartic term cannot be produced through a perturbative calculation. It is nevertheless required by the conservation of the vector current in order to guarantee the invariance of the vector gauge symmetry (which plays a role analogous to the diffeomorphisms in solving the above puzzle).

\section{Conclusions}

The calculations for the odd-parity trace anomaly have led to controversial results, both with pertubative and non-perturbative methods. But while the non-perturbative approaches, if correctly employed, lead to unambiguous results, \cite{bonora2018,Liu2022}, and some disagreements can be attributed to inappropriate methods of calculation (see \cite{bonorasoldati2019} for a discussion), the perturbative ones are intrinsically ambiguous for the reason explained in the previous section. These encompass the perturbative derivations in \cite{bonora2014,bonora2015,bonora2017,Frob2021,Liu2022}. 
As explained before the derivation of a trace anomaly is more complicated than the derivation of the more familiar chiral anomalies and involves the resolution of several ambiguities. The first ambiguity 
is related to the divergent integrals in the Feynman diagrams: it is resolved by a choice of regularization scheme. The second ambiguity lies in the very definition of the trace anomaly and is resolved by formula \eqref{Duff}: as explained in \cite{bonora2021} this formula selects the very violation of the equation of motion while excluding other irrelevant contributions. The third ambiguity is related to cohomology: there is no such thing as a trace anomaly unrelated to diffeomorphisms and other symmetries of the theory. A trace anomaly is a cocycle of the full symmetry of the theory. When we compute a trace anomaly we must make sure that no other symmetry is violated.  Any mis-resolution of these ambiguities may lead to wrong results. For instance, if we compute only the first term in the definition \eqref{Duff} the odd parity trace anomaly disappears. Another example: it is always possible to find a counterterm that completely cancels the lowest order odd-parity trace anomaly, but it inevitably breaks diffeomorphism invariance.

The calculations in \cite{bonora2014,bonora2015,bonora2017,Liu2022} were made by resolving such ambiguities. But, as far as the odd-parity trace anomaly is concerned, there is a fourth ambiguity generated by the vanishing of the odd three-point amplitude of the e.m. tensor. As shown above this implies a dependence of the end result on the regularization scheme. As we have pointed out above, this ambiguity cannot be resolved within the lowest perturbative order, so that this problem is undecidable without going to higher orders of approximation or resorting to non-perturbative methods. Why the perturbative derivations of  \cite{bonora2014,bonora2015,bonora2017,Liu2022} lead anyhow to the correct result is still to be explained.

At this point better avoid any misunderstanding. The true trace anomaly is given by \eqref{Pontryagintrace}. The aim of this note is to point out only the ambiguity of its lowest order perturbative derivation. Another example of the type considered before is related to the odd parity trace anomaly induced by a gauge field, a case recently re-examined in \cite{Bast2022}. This is because the odd part of correlators $<TJJ>$ vanishes identically, just like the odd part of $<TTT>$. In this case there is no need to restrict the cohomology and, anyhow, an appropriate, and quite simple, non-perturbative approach unambiguously leads to a non-vanishing gauge induced trace anomaly, see \cite{bonora2021}. But since in the literature on odd-parity trace anomalies is not univocal, it is worth pointing out that beside the explicit calculations there are also qualitative arguments. To emd this note we would like to briefly recall them. The first is based on the family's index theorem, \cite{atiyahsinger1984}. This theorem can be thought of as expressing the obstructions to the existence of the inverse of the kinetic Dirac-Weyl operator. Any variation of the path integral of the theory defined by the action \eqref{action}, for instance in order to see its response under a Weyl transformation, inevitably involves such an inverse, i.e. the (full)  fermion propagator. The relevant obstructions are expressed in terms of cohomological classes of the classifying space. Among them there are classes that give rise to the well-known chiral consistent anomalies, but in 4d there are also the Pontryagin and Chern classes. The latter are naturally associated to the trace anomaly, generated by the coupling to a background metric or a background gauge field, respectively. 
The second argument is more `phenomenological': the Pontryagin class or the Chern class densities have the right properties and quantum numbers to couple to trace of the e.m. of a system such as \eqref{action}, which violates parity. The experience teaches us that in such cases quantization usually excites such terms (with non vanishing coefficients). The only exceptions may come from (vector) gauge invariance and diffeomorphism invariance. But in this case the latter is satisfied with a non-vanishing Pontryagin term. So the pertinent question would rather be: why should it vanish?

\vskip 1cm
{\bf Note added}. A more comprehensive exposition of the material discussed in this paper can be found in the book \cite{bonora2022}.

\end{document}